\documentclass{jfm}
\usepackage[T1]{fontenc}
\usepackage[latin9]{inputenc}
\usepackage{soul}
\usepackage{xcolor}
\usepackage{graphicx}
\usepackage{amssymb}
\usepackage[authoryear]{natbib}

\makeatletter

\providecommand{\tabularnewline}{\\}
\providecolor{lyxadded}{rgb}{0,0,1}
\providecolor{lyxdeleted}{rgb}{1,0,0}

\ifCUPmtlplainloaded \else
  \IfFileExists{amsbsy.sty}
    {\typeout{^^JFound the 'amsbsy' package on the system, using it.^^J}
}

\def\vec#1{\mbox{\boldmath ${#1}$}}
\def\breve#1{\mbox{\boldmath $\mathsf{#1}$}}
\def\Ca{\mathit{Ca}}
\def\Cm{\mathit{Cm}}
\def\Ha{\mathit{Ha}}

\affiliation{Applied Mathematics Research Centre, Coventry University \\ Priory Street, Coventry CV1 5FB, UK}
\pubyear{2010}
\volume{0}

\begin{document}

\title[Oscillations of weakly viscous conducting liquid drops in a strong magnetic field]
{Oscillations of weakly viscous conducting liquid drops in a strong magnetic field}

\author[J. Priede]{J\ls \=A\ls N\ls I\ls S\ns P\ls R\ls I\ls E\ls D\ls E}

\maketitle
\begin{abstract}
We analyse small-amplitude oscillations of a weakly viscous electrically
conducting liquid drop in a strong uniform DC magnetic field. An asymptotic
solution is obtained showing that the magnetic field does not affect
the shape eigenmodes, which remain the spherical harmonics as in the
non-magnetic case. Strong magnetic field, however, constrains the
liquid flow associated with the oscillations and, thus, reduces the
oscillation frequencies by increasing effective inertia of the liquid.
In such a field, liquid oscillates in a two-dimensional (2D) way as
solid columns aligned with the field. Two types of oscillations are
possible: longitudinal and transversal to the field. Such oscillations
are weakly damped by a strong magnetic field -- the stronger the field,
the weaker the damping, except for the axisymmetric transversal and
inherently 2D modes. The former are overdamped because of being incompatible
with the incompressibility constraint, whereas the latter are not
affected at all because of being naturally invariant along the field.
Since the magnetic damping for all other modes decreases inversely
with the square of the field strength, viscous damping may become
important in a sufficiently strong magnetic field. The viscous damping
is found analytically by a simple energy dissipation approach which
is shown for the longitudinal modes to be equivalent to a much more
complicated eigenvalue perturbation technique. This study provides
a theoretical basis for the development of new measurement methods
of surface tension, viscosity and the electrical conductivity of liquid
metals using the oscillating drop technique in a strong superimposed
DC magnetic field.
\end{abstract}

\section{Introduction}

Shape oscillations of levitated metal droplets can be used to measure
the surface tension and viscosity of liquid metals (\citealt{Rhim-etal99,Egry-etal05}).
Theoretically, the former determines the frequency, while the latter
accounts for the damping rate of oscillations. In the reality, experimental
measurements may be affected by several side-effects. Firstly, levitated
drops may be significantly aspherical and the oscillations amplitudes
not necessarily small, whereas the classical theories describing the
oscillation frequencies (\citealt{Rayl45}) and damping rates (\citealt{Lamb93,Cha81,Reid60})
assume small-amplitude oscillations about an ideally spherical equilibrium
shape. Corrections due to the drop asphericity have been calculated
by \citet{CumBlack91} and \citet{SurBay91}. \citet{BraEgr95} find
the same order correction to the damping rate resulting also from
AC-magnetic field. The effect of a moderate amplitude on the oscillations
of inviscid drops has been analysed by \citet{TsaBro83} who find
that the oscillation frequency decreases with the square of the amplitude.
Using a boundary-integral method, \citet{LunMan91} show that small
viscosity has a relatively large effect on the resonant-mode coupling
phenomena in the nonlinear oscillations of large axially symmetric
drops in zero gravity. Numerical simulation of large-amplitude axisymmetric
oscillations of viscous liquid drop by \citet{Bas92}, who uses the
Galerkin/finite-element technique, shows that a finite viscosity results
in a much stronger mode coupling than predicted by the small-viscosity
approximation.

Secondly, the measurements may strongly be disturbed by AC-driven
flow in the drop. The mode coupling by the internal circulation in
axisymmetrically oscillating drop has been studied numerically by
\citet{MasAsh98} using the Galerkin/finite-element technique. To
reduce the strength of the AC field necessary for the levitation and,
thus, to minimise the flow, experiments may be conducted under the
microgravity conditions during parabolic flights or on the board of
space station (\citealt{Egry-etal99}). A cheaper alternative might
be to apply a sufficiently strong DC magnetic field that can not only
stabilise AC-driven flow but also suppress the convective heat and
momentum transport responsible for the mode coupling under the terrestrial
conditions as originally shown by \citet{SPG03}. Such an approach
has been implemented first by \citet{Yas-etal04} on the electromagnetically
levitated drops of Copper and Nickel which were submitted to a DC
field of the induction up to $10\, T.$ The only motion of $Cu$ drops
observed to persist in magnetic field above $1\, T$ was a solid-body
rotation about an axis parallel to the magnetic field. No shape oscillations,
usually induced by the AC-driven flow fluctuations, were observed.
Note that this implies only the suppression of AC-driven flow but
not of the shape oscillations themselves which require an external
excitation to be observable. \citet{Yas-etal05} study the effect
of suppression of the melt flow on the structure of various alloys
obtained by the electromagnetic levitation melting technique in a
strong superimposed DC magnetic field. The use of high magnetic fields
in various material processing applications is reviewed by \citet{Yas07}. 

Note that a similar suppression of AC-driven flow can also be achieved
by a fast spinning of the drop (\citealt{SPG07}) that may be driven
by an electromagnetic spin-up instability (\citealt{PG00,PG06}).
The effects of both the drop rotation and AC-driven flow on the frequency
spectrum of shape oscillations have been modelled numerically by \citet{BojPer09}.
\citet{Wat09} demonstrates numerically that a large enough oscillation
amplitude can compensate for the effect of rotation on the frequency
shift.

A novel method of measuring thermal conductivity of liquid silicon
using the electromagnetic levitation in a strong superimposed DC magnetic
has been introduced by \citet{Kob-etal07}. Subsequent numerical modelling
by \citet{Tsu-etal09} shows that applying a DC magnetic field of
$4\, T$ can suppress convection in molten silicon droplet enough
to measure its real thermal conductivity. Later on this method has
been extended to the measurements of heat capacity of molten austenitic
stainless steel (\citealt{Fuk-etal09}) and also that of supercooled
liquid silicon (\citealt{Kob-etal10}). 

In order to determine the surface tension and viscosity or the electrical
conductivity one needs to relate the observed surface oscillations
with the relevant thermophysical properties of the liquid. General
small-amplitude shape oscillations of conducting drop in a uniform
DC magnetic field have been analysed first by \citet{Gail66}. Although
a magnetic field of arbitrary strength is considered, the solution
is restricted to inviscid drops. Moreover, only the frequency spectrum
and magnetic damping rates are found but not the associated shape
eigenmodes, which may be useful for experimental identification of
the oscillation modes. Energy dissipation by axisymmetric oscillations
of a conducting drop in a weak DC magnetic field is considered by
\citet{Zamb66}, who finds the magnetic damping rates in agreement
with more general results of \citet{Gail66}. Axisymmetric oscillations
of an electromagnetically levitated drop of molten $Al$ in a superimposed
DC magnetic field are modelled numerically by \citet{BojPer03}. A
moderate DC magnetic field is shown to stabilise AC-driven flow and,
thus, to eliminate the associated shape oscillations. A three-dimensional
numerical simulation of an oscillating liquid metal drop in a uniform
static magnetic field has been carried out by \citet{Tag07}. The
numerical results show that vertical magnetic field effectively damps
the flow, while horizontal field tries to render the flow two-dimensional. 

In the present paper, we analyse free oscillations of a viscous electrically
conducting drop in a homogeneous DC magnetic field. In contrast to
\citet{Gail66}, we assume the viscosity to be small but non-zero
and the magnetic field to be strong. This allows us to obtain an asymptotic
solution to the eigenvalue problem for general small-amplitude 3D
shape oscillations including the eigenmodes left out by \citet{Gail66},
which are necessary for the subsequent determination of the viscous
damping. Firstly, we show that the eigenmodes of shape oscillations
are not affected by strong magnetic field. Namely, they remain the
spherical harmonics as in the non-magnetic case. The magnetic field,
however, changes the internal flow associated with the surface oscillations
and, thus, the frequency spectrum. As the drop oscillates in a strong
magnetic field, the liquid moves as solid columns aligned with the
field. Two types of such oscillations are possible: longitudinal and
transversal to the magnetic field. The oscillations are weakly damped
by a strong magnetic field, except for both the axisymmetric transversal
and inherently 2D modes. The former are magnetically overdamped because
the incompressibility constraint does not permit an axially uniform
radial flow. The latter, which are transversal modes defined by the
spherical harmonics with equal degree and order, $l=m$, are not affected
at all because these modes are naturally invariant along the field.
Because the magnetic damping for all other modes decreases inversely
with the square of the field strength, the viscous damping may become
important in a sufficiently strong magnetic field. 

The paper is organised as follows. The problem is formulated in $\S$\ref{sec:prob}.
Section \ref{sec:invisc} presents an inviscid asymptotic solution
which yields the shape eigenmodes and frequency spectrum of longitudinal
and transversal oscillations. Magnetic damping is found in $\S$\ref{sub:mag-damp}
as a next-order asymptotic correction to the frequency. Viscous damping
rates are calculated in $\S$\ref{sub:visc-damp} first by the eigenvalue
perturbation technique for the longitudinal modes and then by an energy
dissipation approach for both of the oscillation modes. The paper
is concluded by a summary and discussion of the results in $\S$\ref{sec:conc}.

\section{\label{sec:prob}Problem formulation}

\begin{figure}
\begin{centering}
\includegraphics[width=0.33\textwidth]{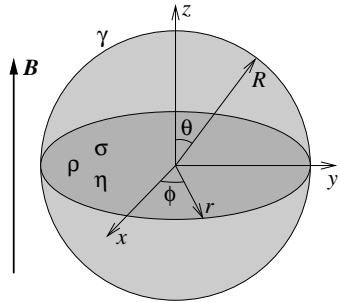}
\par\end{centering}

\caption{\label{fig:Sketch}Sketch to the formulation of problem.}

\end{figure}

Consider a spherical non-magnetic drop of an incompressible liquid
with radius $R_{0},$ density $\rho,$ surface tension $\gamma,$
electrical conductivity $\sigma,$ and a small dynamic viscosity $\eta$
performing small-amplitude shape oscillations in a strong uniform
DC magnetic field $\vec{B}$ as illustrated in figure \ref{fig:Sketch}.
The velocity of liquid flow $\vec{v}$ and the pressure distribution
$p$ are governed by the Navier-Stokes equation with electromagnetic
body force\begin{equation}
\rho\partial_{t}\vec{v}=-\vec{\nabla}p+\eta\vec{\nabla}^{2}\vec{v}+\vec{j}\times\vec{B},\label{eq:NS}\end{equation}
where the induced current follows from Ohm's law for a moving medium
\begin{equation}
\vec{j}=\sigma(\vec{E}+\vec{v}\times\vec{B}).\label{eq:Ohm}\end{equation}
Owing to the smallness of oscillation amplitude, the nonlinear term
in (\ref{eq:NS}) as well as the induced magnetic field are both negligible.
In addition, the characteristic oscillation period $\tau_{0}$ is
supposed to be much longer than the magnetic diffusion time $\mu_{0}\sigma R_{0}^{2},$
where $\mu_{0}$ is the permeability of vacuum. This leads to the
quasi-stationary approximation according to which $\vec{\nabla}\times\vec{E}=0$
and $\vec{E}=-\vec{\nabla}\varphi,$ where $\varphi$ is the electric
potential. The incompressibility constraint $\vec{\nabla}\cdot\vec{v}=0$
and the charge conservation condition $\vec{\nabla}\cdot\vec{j}=0$
applied to (\ref{eq:NS}) and (\ref{eq:Ohm}) result, respectively,
in\begin{eqnarray}
\vec{\nabla}^{2}p & = & \sigma(\vec{B}\cdot\vec{\nabla})(\vec{B}\cdot\vec{v}),\label{eq:p}\\
\vec{\nabla}^{2}\varphi & = & \vec{B}\cdot\vec{\nabla}\times\vec{v}.\label{eq:phi}\end{eqnarray}
For a uniform $\vec{B}$ under consideration here, applying the operators
$\vec{\nabla}\times\vec{\nabla}\times,$ $(\vec{B}\cdot\vec{\nabla})\vec{B}\cdot$
and $(\vec{B}\cdot\vec{\nabla})\vec{B}\cdot\vec{\nabla}\times$ to
(\ref{eq:NS}) and taking into account $\vec{\nabla}\times\vec{\nabla}\times(\vec{j}\times\vec{B})=\sigma(\vec{B}\cdot\vec{\nabla})^{2}\vec{v}$
together with (\ref{eq:p}) and (\ref{eq:phi}), we obtain\begin{equation}
\left[\rho\vec{\nabla}^{2}\partial_{t}+\sigma(\vec{B}\cdot\vec{\nabla})^{2}-\eta\vec{\nabla}^{4}\right]\{p,\varphi,\vec{v}\}=0.\label{eq:pfv}\end{equation}
Although the equation above applies to $p,$ $\varphi$ and $\vec{v}$
separately, these variables are not independent of each other. Firstly,
owing to the incompressibility constraint, only two velocity components
are mutually independent. Secondly, velocity is related to the pressure
and electric potential by (\ref{eq:NS}), which can be used to represent
$\vec{v}$ in terms of $p$ and $\varphi$ as done in the following.

Boundary conditions are applied at the drop surface $S$ defined by
its spherical radius $R=R_{0}+R_{1}(\theta,\phi,t),$ where $R_{1}$
is a small perturbation, which depends on the poloidal and azimuthal
angles, $\theta$ and $\phi,$ and the time $t$. The radial velocity
at the surface is related to the radius perturbation by the kinematic
constraint \begin{equation}
\left.v_{R}\right|_{S}=\partial_{t}R_{1}.\label{eq:kinc}\end{equation}
 The normal component of the current at the drop surface, which is
assumed to be surrounded by vacuum or insulating gas, vanishes, i.e.,
$\left.j_{n}\right|_{S}=0.$ In addition, there is no tangential stress
at the free surface: \begin{equation}
\left.\vec{n}\cdot\partial_{\tau}\vec{v}+\vec{\tau}\cdot\partial_{n}\vec{v}\right|_{S}=0,\label{eq:tblnc}\end{equation}
 while the normal stress component is balanced by the capillary pressure
\begin{equation}
p_{0}+p-2\eta\partial_{n}v_{n}=\gamma\vec{\nabla}\cdot\vec{n},\label{eq:nblnc}\end{equation}
where $p_{0}=2\gamma/R_{0}$ is the constant part of pressure, $\vec{\tau}$
is a unit tangent vector and $\vec{n}=\vec{\nabla}(R-R_{1})/|\vec{\nabla}(R-R_{1})|$
is the outward surface normal. For small-amplitude oscillations defined
by $R_{1}\ll R_{0},$ we have $\vec{n}\approx\vec{e}_{R}-\vec{\nabla}R_{1}.$ 

Henceforth, we proceed to dimensionless variables by choosing the
radius $R_{0}$ and the characteristic capillary pressure $P_{0}=\gamma/R_{0}$
as the length and pressure scales. The characteristic period of capillary
oscillations is determined by the balance of inertia and pressure,
which yields the time scale $\tau_{0}=\sqrt{R_{0}^{3}\rho/\gamma}.$
The velocity and potential scales are chosen as $v_{0}=R_{0}/\tau_{0}$
and $\varphi_{0}=v_{0}BR_{0},$ respectively, where $B=\left|\vec{B}\right|.$
In the dimensionless variables, (\ref{eq:pfv}) takes the form\begin{equation}
\left[\vec{\nabla}^{2}\partial_{t}+\Cm(\vec{\epsilon}\cdot\vec{\nabla})^{2}-\Ca\vec{\nabla}^{4}\right]\{p,\varphi,\vec{v}\}=0,\label{eq:pfv-nd}\end{equation}
where $\vec{\epsilon}=\vec{B}/B$ is a unit vector in the direction
of the magnetic field and $\Ca=\eta/\sqrt{R_{0}\rho\gamma}$ and $\Cm=\sigma B^{2}R_{0}^{2}/\sqrt{R_{0}\rho\gamma}$
are the conventional and magnetic capillary numbers, respectively.
They are the ratios of the capillary oscillation time $\tau_{0}$
defined above and the viscous and magnetic damping times, which are
$\tau_{v}=\rho R_{0}^{2}/\eta$ and $\tau_{m}=\rho/(\sigma B^{2}),$
respectively. In the dimensionless form, the normal stress balance
condition (\ref{eq:nblnc}) reads as

\begin{equation}
\left.(\vec{\nabla}^{2}+2)R_{1}+p-2\Ca\partial_{R}v_{R}\right|_{R=1}=0.\label{eq:nblnc-nd}\end{equation}
In the following, we assume viscosity to be small but the magnetic
field strong so that $\Ca\ll1$ and $\Cm\gg1,$ which means that the
second and third terms in (\ref{eq:pfv-nd}) are much greater and
much smaller, respectively, than the first one. Thus, we first focus
on the effect of the magnetic field and ignore that of viscosity,
which is considered later in $\S$\ref{sub:visc-damp}.

\section{\label{sec:invisc}Inviscid asymptotic solution}

Here we ignore viscosity that allows us to formulate the problem in
terms of $p,$ $\varphi$ and $R_{1}.$ Projecting the dimensionless
counterpart of (\ref{eq:NS}), which takes the form \begin{equation}
\Cm\vec{v}+\partial_{t}\vec{v}=-\vec{\nabla}p+\Ca\vec{\nabla}^{2}\vec{v}+\Cm\left[\vec{\epsilon}\times\vec{\nabla}\varphi+\vec{\epsilon}(\vec{\epsilon}\cdot\vec{v})\right],\label{eq:NS-nd}\end{equation}
onto $\vec{e}_{R}$ and $\vec{\epsilon},$ and putting $\Ca=0,$ we
obtain \begin{eqnarray}
\Cm v_{R}+\partial_{t}v_{R} & = & -\vec{e}_{R}\cdot\vec{\nabla}p+\Cm\left[\vec{e}_{R}\times\vec{\epsilon}\cdot\vec{\nabla}\varphi+\vec{e}_{R}\cdot\vec{\epsilon}v_{\shortparallel}\right],\label{eq:NS-R}\\
\partial_{t}v_{\shortparallel} & = & -\vec{\epsilon}\cdot\vec{\nabla}p,\label{eq:NS-||}\end{eqnarray}
where $v_{\shortparallel}=\vec{\epsilon}\cdot\vec{v}$ is the velocity
component along the magnetic field. Differentiating (\ref{eq:NS-R})
with respect to $t$ and substituting $\partial_{t}v_{\shortparallel}$
from (\ref{eq:NS-||}), we represent (\ref{eq:kinc}) in terms of
$p$ and $\varphi$\begin{equation}
\Cm\partial_{t}^{2}R_{1}+\partial_{t}^{3}R_{1}=\left[\Cm\left(\vec{e}_{R}\times\vec{\epsilon}\cdot\vec{\nabla}\partial_{t}\varphi-(\vec{e}_{R}\cdot\vec{\epsilon})\vec{\epsilon}\cdot\vec{\nabla}p\right)-\vec{e}_{R}\cdot\vec{\nabla}\partial_{t}p\right]_{R=1}.\label{eq:R1-cnd}\end{equation}
 Velocity has to be eliminated also from the electric boundary condition
given by the radial component of Ohm's law\begin{equation}
\left.j_{R}\right|_{R=1}=-\vec{e}_{R}\cdot\left[\vec{\nabla}\varphi+\vec{\epsilon}\times\vec{v}\right]_{R=1}=0.\label{eq:jR}\end{equation}
Firstly, applying $(\Cm+\partial_{t})$ to (\ref{eq:jR}) and then
using (\ref{eq:NS-nd}), we obtain \begin{equation}
\left[\Cm(\vec{e}_{R}\cdot\vec{\epsilon})\vec{\epsilon}\cdot\vec{\nabla}\varphi-\vec{e}_{R}\times\vec{\epsilon}\cdot\vec{\nabla}p+\vec{e}_{R}\cdot\vec{\nabla}\partial_{t}\varphi\right]_{R=1}=0.\label{eq:phi-cnd}\end{equation}
In the inviscid approximation, (\ref{eq:nblnc-nd}) reduces to \begin{equation}
\left.p\right|_{R=1}=-(\vec{\nabla}^{2}+2)R_{1}.\label{eq:p-cnd}\end{equation}
In the following, besides the spherical coordinates $(R,\theta,\phi),$
we will be using also the cylindrical ones $(r,\phi,z)$ with the
axis aligned along the magnetic field so that $\vec{\epsilon}=\vec{e}_{z}.$ 

Solution is sought in the normal mode form $\{p,\varphi,R_{1}\}=\{\hat{p},\hat{\varphi},\hat{R}\}(\vec{r})e^{\beta t+im\phi},$
where $\hat{p},$ $\hat{\varphi}$ and $\hat{R}$ are axisymmetric
amplitude distributions, $m$ is the azimuthal wave number, and $\beta$
is a generally complex temporal variation rate which has to be determined
depending on $m,$ $\Cm$ and $\Ca.$ Then boundary conditions (\ref{eq:R1-cnd}),
(\ref{eq:phi-cnd}) and (\ref{eq:p-cnd}) for the oscillation amplitudes
at $R=1$ take the form \begin{eqnarray}
\beta^{2}\hat{R}+im\beta\hat{\varphi}+z\partial_{z}\hat{p} & = & -\Cm^{-1}(\beta^{3}\hat{R}+\beta\partial_{R}\hat{p}),\label{eq:Rh-cnd}\\
z\partial_{z}\hat{\varphi} & = & -\Cm^{-1}(im\hat{p}+\beta\partial_{R}\hat{\varphi}),\label{eq:phih-cnd}\\
\hat{p} & = & -(L_{z}+2)\hat{R},\label{eq:ph-cnd}\end{eqnarray}
where $L_{z}\equiv\frac{d}{dz}\left((1-z^{2})\frac{d\,}{dz}\right)-\frac{m^{2}}{1-z^{2}}$
is the angular part of the Laplace operator in the spherical coordinates
for the azimuthal mode $m$ written in terms of $z=\cos\theta.$ Further,
it is important to note that\begin{equation}
L_{z}P_{l}^{m}(z)=-l(l+1)P_{l}^{m}(z),\label{eq:Legendre}\end{equation}
where $P_{l}^{m}(z),$ the associated Legendre function of degree
$l$ and order $m,$ is an eigenfunction of $L_{z}$ with eigenvalue
$-l(l+1)$ (\citealt{AbSt72}). Equation (\ref{eq:pfv-nd}) for $\hat{p}$
and $\hat{\phi}$ can be written as \begin{equation}
\left[\partial_{z}^{2}+\Cm^{-1}(L_{r}+\partial_{z}^{2})(\beta-\Ca(L_{r}+\partial_{z}^{2}))\right]\{\hat{p},\hat{\varphi}\}=0,\label{eq:phip}\end{equation}
where $L_{r}\equiv\partial_{r}^{2}+r^{-1}\partial_{r}-m^{2}/r^{2}$
is the radial part of the Laplace operator in the cylindrical coordinates
for the azimuthal mode $m.$ Here we put $\Ca=0,$ suppose $\Cm\gg1,$
and search for an asymptotic solution in the terms of a small parameter
$\Cm^{-1}$ as\[
\{\hat{p},\hat{\varphi,}\hat{R},\beta\}\sim\{\hat{p}_{0},\hat{\varphi}_{0},\hat{R}_{0},\beta_{0}\}+\Cm^{-1}\{\hat{p}_{1},\hat{\varphi}_{1},\hat{R}_{1},\beta_{1}\}+\cdots.\]
Note that although (\ref{eq:phip}) admits solutions with $\beta\sim\Cm$
found by \citet{Gail66}, such quickly relaxing modes cannot be related
with the surface deformations. From the physical point of view, drop
is driven to its equilibrium shape by the surface tension, and the
magnetic field can only oppose but not to accelerate the associated
liquid flow. From the mathematical point of view, $\beta\sim\Cm\gg1$
applied to (\ref{eq:Rh-cnd}) results in $\hat{R}_{0}=0,$ which means
no surface deformation at the leading order in agreement with the
previous physical arguments. Consequently, these fast modes represent
internal flow perturbations which are not relevant for the shape deformations
under consideration here.

\subsection{\label{sub:osc-freq}Oscillation frequencies}

At the leading order, (\ref{eq:phip}) reduces to $\partial_{z}^{2}\left\{ \hat{p}_{0},\hat{\varphi}_{0}\right\} =0,$
whose general solution is \begin{equation}
\{\hat{p}_{0},\hat{\varphi}_{0}\}(r,z)=\{\hat{p}_{0}^{+},\hat{\varphi}_{0}^{+}\}(r)+z\{\hat{p}_{0}^{-},\hat{\varphi}_{0}^{-}\}(r),\label{eq:p0f0}\end{equation}
where the first pair of particular solutions are the functions of
$r$ only, while the second pair is linear in $z$ but general in
$r.$ Owing to the $z$-reflection symmetry of the problem these two
types particular of solutions do not mix and, thus, they are subsequently
considered separately. We refer to these solutions in accordance to
their $z$-parity as even and odd ones using the indices $e$ and
$o.$ As shown below, the odd and even solutions describe longitudinal
and transversal oscillation modes, respectively.

\subsubsection{\label{sub:L-freq}Longitudinal modes}

For the odd solutions $\{\hat{p}_{0}^{o},\hat{\varphi}_{0}^{o}\}(r,z)=z\{\hat{p}_{0}^{-},\hat{\varphi}_{0}^{-}\}(r),$
boundary condition (\ref{eq:phih-cnd}), which at the leading order
reads as $z\partial_{z}\hat{\varphi}_{0}=0,$ results in $\hat{\varphi}_{0}^{-}(r)=0.$
The two remaining boundary conditions (\ref{eq:Rh-cnd}) and (\ref{eq:ph-cnd})
take the form\begin{eqnarray}
{\beta_{0}^{o}}^{2}\hat{R}_{0}^{o} & = & -z\hat{p}_{0}^{-},\label{eq:R0-o}\\
(L_{z}+2)\hat{R}_{0}^{o} & = & -z\hat{p}_{0}^{-}.\label{eq:p0-o}\end{eqnarray}
Eliminating the pressure term between the equations above, we obtain
an eigenvalue problem in $\beta_{0}^{2}$ for $\hat{R}_{0}^{o}$ \begin{equation}
(L_{z}+2-{\beta_{0}^{o}}^{2})\hat{R}_{0}^{o}=0,\label{egv:R0-o}\end{equation}
which is easily solved by using (\ref{eq:Legendre}) as\begin{eqnarray}
\hat{R}_{0}^{o}(z) & = & R_{0}^{o}P_{l}^{m}(z),\label{sol:R0-o}\\
\beta_{0}^{o} & = & \pm i\sqrt{(l-1)(l+2)},\label{sol:bt0-o}\end{eqnarray}
where $R_{0}^{o}$ is a small amplitude of oscillations and $l-m$
is an odd positive number. Note that imaginary $\beta_{0}^{o}$ describes
constant-amplitude harmonic oscillations with the circular frequency
$|\beta_{0}^{o}|$ which differs from the corresponding non-magnetic
result only by the factor of $\sqrt{l}$ (\citealt{Lamb93}), and
coincides with the result stated by \citet{Gail66}. Thus, strong
magnetic field changes only the eigenfrequencies but not the eigenmodes
of shape oscillations which, as without the magnetic field, are represented
by separate spherical functions (associated Legendre functions with
integer indices) (\citealt{AbSt72}). Similarly to the non-magnetic
case, the frequency spectrum for odd modes is degenerate because it
depends only on the degree $l$ but not on the order $m.$ Thus, for
each $l,$ there are $[l/2]$ odd modes with different $m.$ 

\begin{figure}
\begin{centering}
\includegraphics[width=0.25\textwidth]{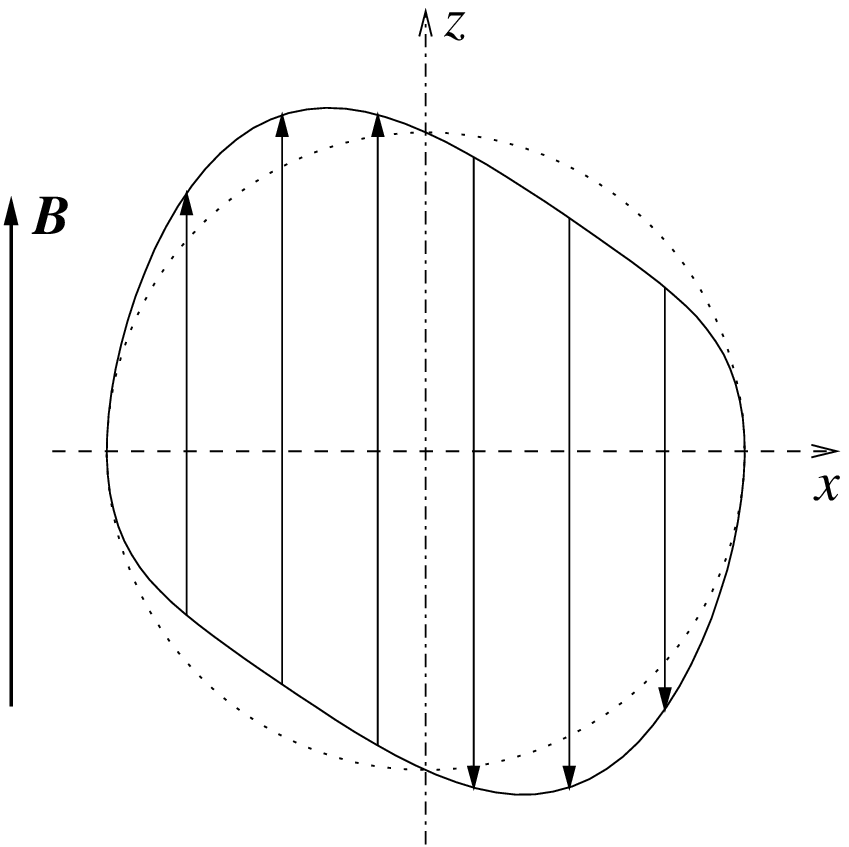}\put(-100,0){(a)}\includegraphics[width=0.25\textwidth]{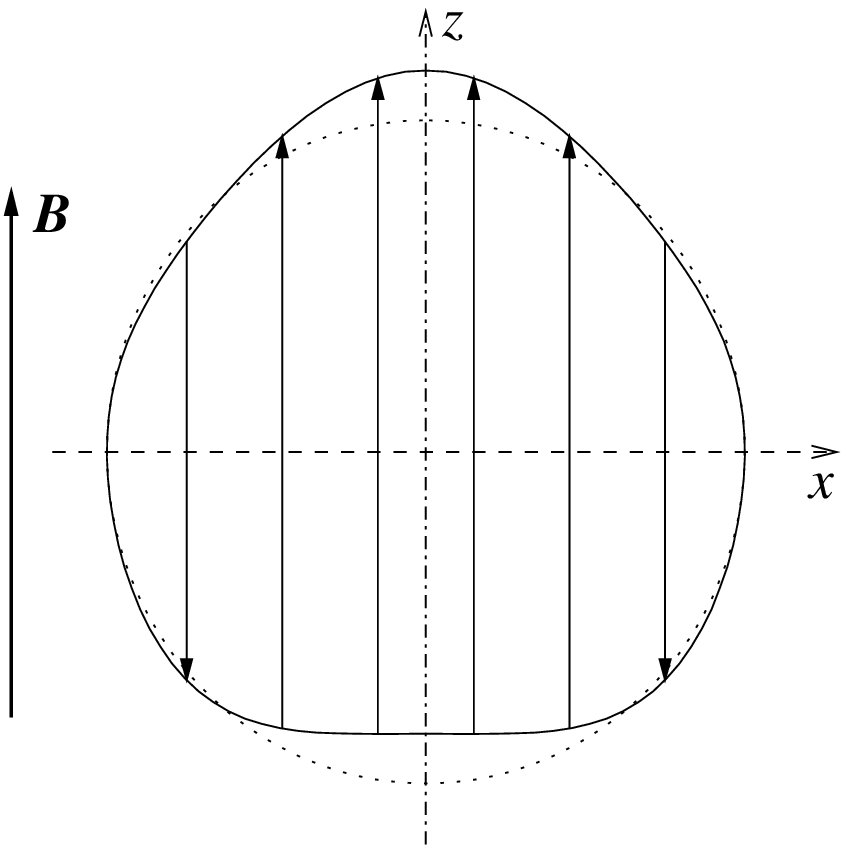}\put(-100,0){(b)}\includegraphics[width=0.25\textwidth]{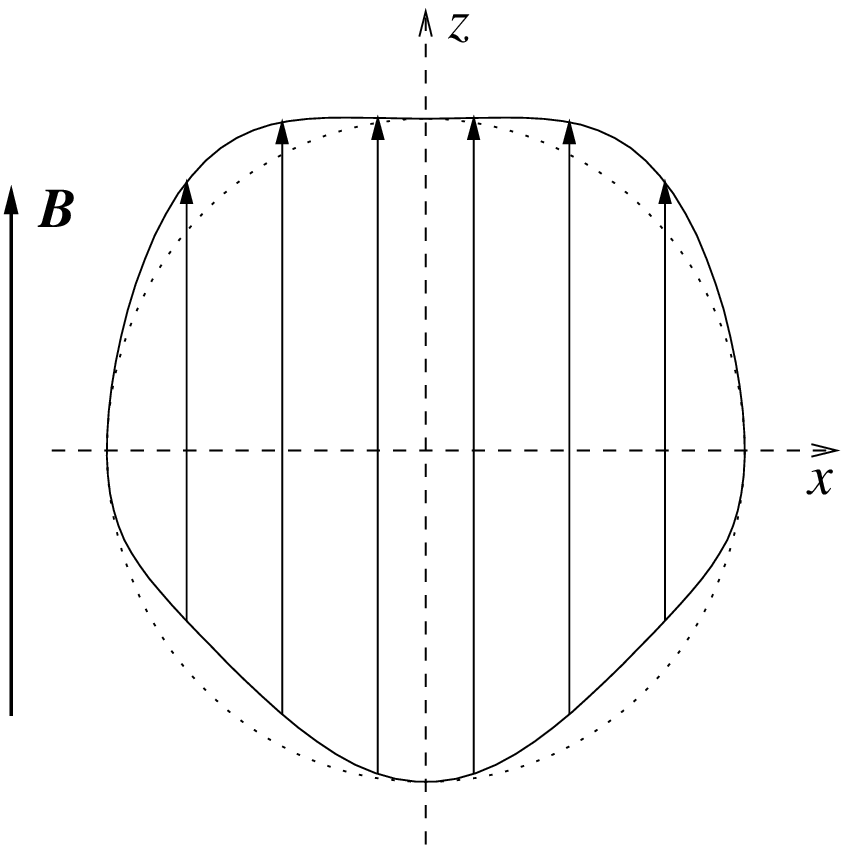}\put(-100,0){(c)}\includegraphics[width=0.25\textwidth]{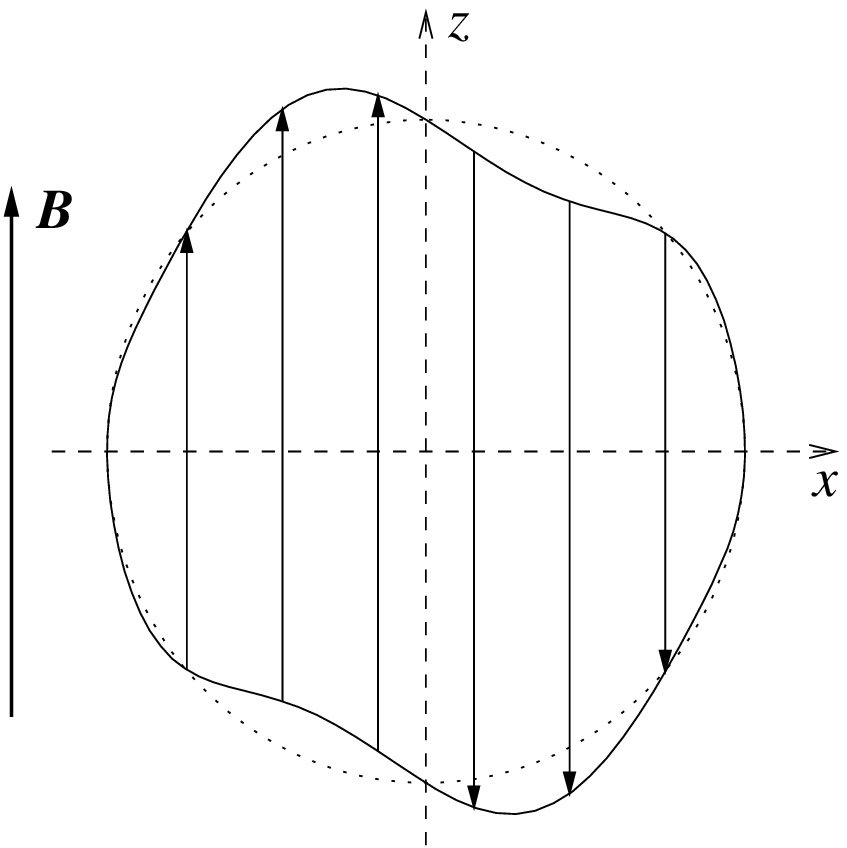}\put(-100,0){(d)}\caption{\label{fig:odd}Shapes and the associated liquid oscillations in the
$(x,z)$-plane parallel to the magnetic field for the first four longitudinal
oscillation modes with indices ($l,m)=(2,1)$ (a), $(3,0)$ (b), $(3,2)$
(c), and (4,1) (d).}

\par\end{centering}
\end{figure}

Taking into account that $z=\sqrt{1-r^{2}}$ at the surface, the radial
pressure distribution is obtained from (\ref{eq:R0-o}) as \begin{equation}
\hat{p}_{0}^{-}(r)=-\beta_{0}^{2}\hat{R}_{0}^{o}(\sqrt{1-r^{2}})/\sqrt{1-r^{2}}.\label{sol:p0-m}\end{equation}
According to (\ref{eq:NS-||}), this pressure distribution is associated
with the axial velocity component \begin{equation}
\hat{w}_{0}^{o}(r)=-\beta_{0}^{-1}\hat{p}_{0}^{-}(r),\label{eq:w0-o}\end{equation}
while two other velocity components transversal to the magnetic field
are absent in the leading-order approximation. Thus, the liquid effectively
oscillates in solid columns along the magnetic field as illustrated
in figure \ref{fig:odd} for the first four longitudinal oscillation
modes defined by the indices $(l,m)=(2,1)$, $(3,0)$, $(3,2),$and
$(4,1).$ Since such a flow does not cross the flux lines, the oscillations
are not damped by the magnetic field in the leading-order approximation. 

\subsubsection{\label{sub:T-freq}Transversal modes}

For the even solutions $\{\hat{p}_{0}^{e},\hat{\varphi}_{0}^{e}\}(r,z)=\{\hat{p}_{0}^{+},\hat{\varphi}_{0}^{+}\}(r),$
the leading-order boundary condition (\ref{eq:phih-cnd}) is satisfied
automatically. The two remaining conditions (\ref{eq:Rh-cnd}) and
(\ref{eq:ph-cnd}) then take the form\begin{eqnarray}
\beta_{0}^{e}\hat{R}_{0}^{e}+im\hat{\varphi}_{0}^{+} & = & 0,\label{eq:R0-e}\\
(L_{z}+2)\hat{R}_{0}^{e} & = & -\hat{p}_{0}^{+}.\label{eq:p0-e}\end{eqnarray}
In contrast to the longitudinal modes considered above, now we have
two equations (\ref{eq:R0-e}) and (\ref{eq:p0-e}) but three unknowns.
To solve this problem, we need to consider the first-order solution
to (\ref{eq:phip}) which now takes the form $\partial_{z}^{2}\{\hat{p}_{1}^{e},\hat{\varphi}_{1}^{e}\}=-\beta_{0}L_{r}\{\hat{p}_{0}^{+},\hat{\varphi}_{0}^{+}\}$
and yields \begin{equation}
\{\hat{p}_{1}^{e},\hat{\varphi}_{1}^{e}\}(r,z)=\{\hat{p}_{1}^{+},\hat{\varphi}_{1}^{+}\}(r)-\frac{1}{2}\beta_{0}^{e}z^{2}L_{r}\{\hat{p}_{0}^{+},\hat{\varphi}_{0}^{+}\}.\label{eq:p-phi-1e}\end{equation}
Then boundary condition (\ref{eq:phih-cnd}) results in $imp_{0}^{+}-\beta_{0}^{e}(z^{2}L_{r}-r\partial_{r})\hat{\varphi}_{0}^{+}=0.$
Combining this with (\ref{eq:R0-e}) and (\ref{eq:p0-e}) and taking
into account \begin{equation}
\left.z^{2}L_{r}-r\partial_{r}\right|_{R=1}\equiv L_{z}+m^{2},\label{eq:Lr-Lz}\end{equation}
we obtain \begin{equation}
\left[L_{z}+2+(\beta_{0}^{e}/m)^{2}(L_{z}+m^{2})\right]\hat{R}_{0}^{e}=0.\label{egv:R0-e}\end{equation}
Using (\ref{eq:Legendre}), we readily obtain \begin{eqnarray}
\hat{R}_{0}^{e}(z) & = & R_{0}^{e}P_{l}^{m}(z),\label{sol:R0-e}\\
\beta_{0}^{e} & = & \pm im\sqrt{\frac{(l-1)(l+2)}{l(l+1)-m^{2}}},\label{sol:bt0-e}\end{eqnarray}
where $R_{0}^{e}$ is a small oscillation amplitude and $l-m$ is
an even non-negative number. The result above again agrees with the
asymptotic solution given by \citet{Gail66}. Similarly to the odd
solutions found in the previous section, even eigenmodes are represented
by separate spherical functions, and the oscillations are not damped
at the leading order. In contrast to the odd modes as well as to the
non-magnetic case, the frequency spectrum (\ref{sol:bt0-e}) is no
longer degenerate and frequencies vary with the azimuthal wave number
$m$. In particular, there are two important results implied by (\ref{sol:bt0-e}).
Firstly, the oscillation frequency for the axisymmetric modes specified
by $m=0$ is zero. This means that these modes are over-damped and
do not oscillate at all. Secondly, the oscillation frequency for the
modes with $m=l$ is exactly the same as without the magnetic field,
i.e., $\sqrt{l(l-1)(l+2)}.$ This is because the liquid flow associated
with these oscillation modes is inherently invariant along the field
and, thus, not affected by the last (\citealt{Gail66}).

The electric potential and pressure distributions follow from (\ref{eq:R0-e})
and (\ref{eq:p0-e}) as \begin{eqnarray}
\hat{\varphi}_{0}^{e}(r) & = & im^{-1}\beta_{0}^{e}\hat{R}_{0}^{e}(\sqrt{1-r^{2}}),\label{sol:phi0-p}\\
\hat{p}_{0}^{e}(r) & = & (l-1)(l+2)\hat{R}_{0}^{e}(\sqrt{1-r^{2}}).\label{sol:p0-p}\end{eqnarray}
The associated velocity distribution is obtained from (\ref{eq:NS-nd}).
Firstly, equation (\ref{eq:NS-||}) implies that the liquid oscillations
are purely transversal to the magnetic field. In the leading-order
terms, we obtain from (\ref{eq:NS-nd}) \begin{equation}
\vec{v}_{0}^{e}(r,\phi)=\vec{e}_{z}\times\vec{\nabla}\varphi_{0}^{e}(r,\phi),\label{sol:v0-e}\end{equation}
which shows that the velocity is not only transversal but also invariant
along the magnetic field. Thus, the liquid again oscillates as solid
columns, but in this case transversely to the field which has no effect
on such a flow. This is because the e.m.f induced by the flow, which
is invariant along the magnetic field, is irrotational, i.e., $\vec{\nabla}\times(\vec{v}\times\vec{B})=(\vec{B}\cdot\vec{\nabla})\vec{v}\equiv0,$
and, thus, unable to drive current circulation in a closed liquid
volume. 

Note that for the axisymmetric modes $(m=0),$ the potential (\ref{sol:phi0-p})
and the associated velocity (\ref{sol:v0-e}) take an indeterminate
form. Namely, for $m=0,$ boundary condition (\ref{eq:R0-e}), which
in this is case straightforwardly implies a zero frequency, is satisfied
by an arbitrary potential distribution independent of the radius perturbation.
As seen from (\ref{sol:phi0-p}), a non-zero axisymmetric potential
is associated with a purely azimuthal velocity. Consequently, this
mode is irrelevant and can subsequently be neglected because it represents
an internal flow perturbation which is just compatible but not coupled
with axisymmetric shape deformations similarly to the fast modes discussed
at the end of $\S$\ref{sec:invisc}. Moreover, this is consistent
with (\ref{sol:bt0-e}) according to which axisymmetric transversal
modes are static in the leading-order approximation that implies a
zero velocity and, consequently, a zero associated potential. 

\begin{figure}
\begin{centering}
\includegraphics[width=0.25\textwidth]{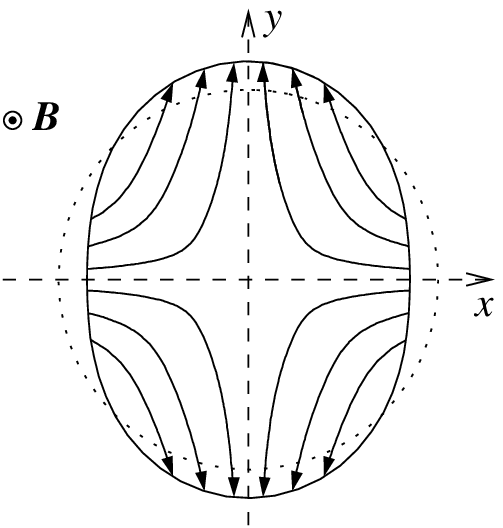}\put(-110,0){(a)}\includegraphics[width=0.25\textwidth]{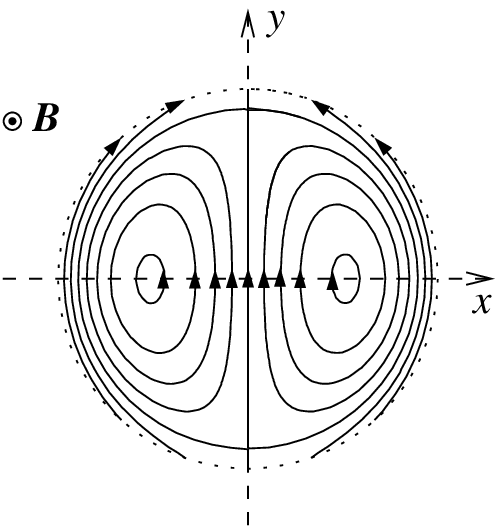}\put(-110,0){(b)}\includegraphics[width=0.25\textwidth]{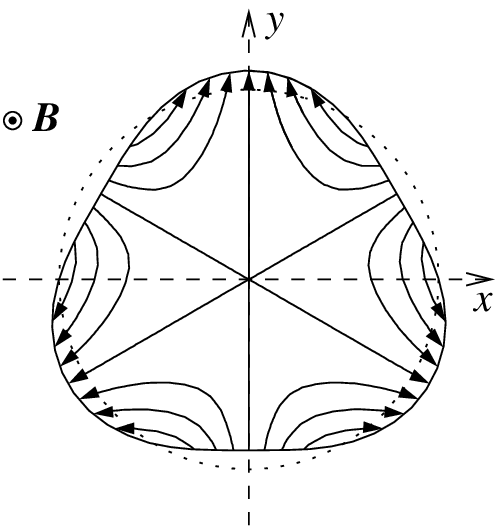}\put(-110,0){(c)}\includegraphics[width=0.25\textwidth]{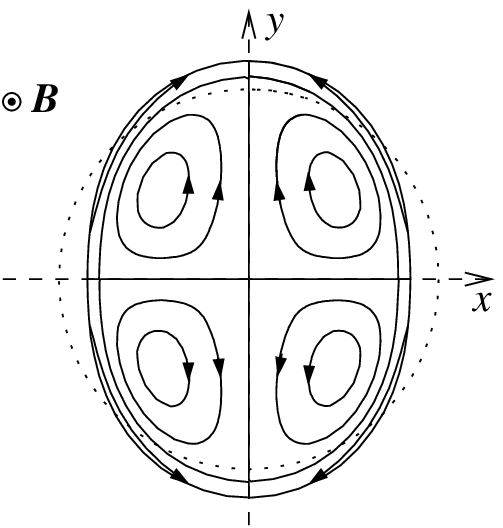}\put(-110,0){(d)}\caption{\label{fig:even}Shapes and the associated liquid flows in the horizontal
mid-plane $(z=0)$ perpendicular to the magnetic field for the first
four transversal oscillation modes defined by indices $(l,m)=(2,2)$
(a), $(3,1)$ (b), (3,3) (c) and $(4,2)$ (d).}

\par\end{centering}
\end{figure}

Expression (\ref{sol:v0-e}) implies that the velocity streamlines
coincide with the isolines of $\varphi_{0},$ which, thus, represents
a stream function for the flow oscillations. Figure \ref{fig:even}
shows the shapes and streamlines of the associated liquid flow in
the horizontal mid-plane for the first four transversal oscillation
modes. Note that the first and the third mode with the indices $(l,m)=(2,2)$
and $(3,3)$ are both naturally invariant in the direction of the
magnetic field and, thus, effectively non-magnetic. The second mode
with $(l,m)=(3,1)$ corresponds to the drop oscillating in such a
way that horizontal cross-sections remain circular in the small-amplitude
limit under consideration while the whole shape deforms because of
vertical offset of the cross-sections. 

\subsection{\label{sub:mag-damp}Magnetic damping}

\subsubsection{Longitudinal modes}

In order to determine the magnetic damping rates for longitudinal
modes, we have to consider the first-order solution governed by\[
\partial_{z}^{2}\{\hat{p}_{1}^{o},\hat{\varphi}_{1}^{o}\}=-\beta_{0}^{o}zL_{r}\{\hat{p}_{0}^{-},0\},\]
which yields\[
\{\hat{p}_{1}^{o},\hat{\varphi}_{1}^{o}\}(r,z)=z\{\hat{p}_{1}^{-},\hat{\varphi}_{1}^{-}\}(r)-\frac{1}{6}\beta_{0}^{o}z^{3}L_{r}\{\hat{p}_{0}^{-},0\}.\]
Then (\ref{eq:phih-cnd}) and (\ref{eq:R0-o}) applied consecutively
result in $z\hat{\varphi}_{1}^{-}=-imz\hat{p}_{0}^{-},$ which combined
with (\ref{eq:Rh-cnd}), (\ref{eq:ph-cnd}) and (\ref{eq:R0-o}) yields
\[
(L_{z}+2-{\beta_{0}^{o}}^{2})\hat{R}_{1}^{o}=\frac{1}{3}{\beta_{0}^{o}}^{3}z\left[z^{2}L_{r}-3r\partial_{r}\right]z^{-1}\hat{R}_{0}^{o}+\beta_{0}^{o}(2\beta_{1}^{o}-m^{2}{\beta_{0}^{o}}^{2})\hat{R}_{0}^{o}.\]
After some algebra, we obtain $\left.z\left[z^{2}L_{r}-3r\partial_{r}\right]z^{-1}\right|_{R=1}\equiv L_{z}+2+m^{2},$
and, consequently, \begin{equation}
(L_{z}+2-{\beta_{0}^{o}}^{2})\hat{R}_{1}^{o}=\frac{1}{3}\beta_{0}^{o}\left[{\beta_{0}^{o}}^{2}(L_{z}-2m^{2}+2)+6\beta_{1}^{o}\right]\hat{R}_{0}^{o}.\label{egv:R1-o}\end{equation}
The l.h.s. operator above is the same as that in (\ref{egv:R0-o})
which has $\hat{R}_{0}^{o}$ as its eigensolution with a zero eigenvalue.
Owing to (\ref{eq:Legendre}) and (\ref{sol:R0-o}), $\hat{R}_{0}^{o}$
is an eigensolution of the r.h.s operator of (\ref{egv:R1-o}), too.
Thus, for (\ref{egv:R1-o}) to be solvable, its r.h.s has to be free
of the terms proportional to $\hat{R}_{0}^{o},$ that yields \begin{equation}
\beta_{1}^{o}=-\frac{1}{6}(l-1)(l+2)((l-1)(l+2)+2m^{2}).\label{eq:bt1-o}\end{equation}
Note that conversely to the frequency for longitudinal oscillation
modes (\ref{sol:bt0-o}), the magnetic damping rate above is not degenerate
and varies with $m.$

\subsubsection{Transversal modes}

Similarly to the oscillation frequency considered above, boundary
conditions (\ref{eq:ph-cnd}) and (\ref{eq:Rh-cnd}) applied to the
first-order solution (\ref{eq:p-phi-1e}) result in \begin{eqnarray}
\beta_{0}\hat{R}_{1}^{e}+im\hat{\varphi}_{1}^{+} & = & \left[{\beta_{0}^{e}}^{2}\left(m^{-2}(z^{2}L_{r}-r\partial_{r})^{2}-\frac{1}{2}z^{2}L_{r}-1\right)-\beta_{1}^{e}\right]\hat{R}_{0}^{e},\label{eq:R1-e}\\
(L_{z}+2)\hat{R}_{1}^{e} & = & -\hat{p}_{1}^{e}.\label{eq:p1-e}\end{eqnarray}
To solve this first-order problem, we again need a second-order solution
governed by\[
\partial_{z}^{2}\{\hat{p}_{2}^{e},\hat{\varphi}_{2}^{e}\}=-\beta_{0}^{e}(L_{r}+\partial_{z}^{2})\{\hat{p}_{1}^{e},\hat{\varphi}_{1}^{e}\}-\beta_{1}^{e}L_{r}\{\hat{p}_{0}^{+},\hat{\varphi}_{0}^{+}\},\]
which, by taking into account (\ref{eq:p-phi-1e}), yields \begin{equation}
\hat{\varphi}_{2}^{e}(r,z)=\hat{\varphi}_{2}^{+}(r)-\frac{1}{2}\beta_{0}^{e}z^{2}L_{r}\hat{\varphi}_{1}^{+}+{\beta_{0}^{e}}^{2}\left[\frac{z^{2}}{2}L_{r}+\frac{z^{4}}{4!}L_{r}^{2}\right]\hat{\varphi}_{0}^{+}-\frac{1}{2}\beta_{1}^{e}z^{2}L_{r}\hat{\varphi}_{0}^{+},\label{eq:p-phi-2e}\end{equation}
Then (\ref{eq:ph-cnd}) results in\begin{equation}
im\hat{p}_{1}^{e}-\beta_{0}^{e}(z^{2}L_{r}-r\partial_{r})\hat{\varphi}_{1}^{+}=im^{-1}\left[\beta_{0}^{e}(z^{2}L_{r}-r\partial_{r})-\frac{1}{6}{\beta_{0}^{e}}^{2}z^{2}(z^{2}L_{r}-3r\partial_{r})L_{r}\right]\hat{R}_{0}^{e}.\label{eq:phi2-e}\end{equation}
Substituting $\hat{\varphi}_{1}^{+}$ and $\hat{p}_{1}^{e}$ from
(\ref{eq:R1-e}) and (\ref{eq:phi2-e}) into (\ref{eq:p1-e}) and
using \[
\left.z^{2}(z^{2}L_{r}-3r\partial_{r})L_{r}-3(z^{2}L_{r}-r\partial_{r})z^{2}L_{r}\right|_{R=1}\equiv2m^{2}-2(L_{z}+m^{2})^{2},\]
after some algebra we obtain an equation for $\hat{R}_{1}^{e},$ which
is the same as (\ref{egv:R0-e}) for $\hat{R}_{0}^{e},$ except for
the r.h.s. that now reads as

\[
\frac{\beta_{0}^{e}}{3m^{2}}\left[(\beta_{0}^{e}/m)^{2}\left((L_{z}+m^{2})^{2}-m^{2}\right)(3L_{z}+2m^{2})-6\beta_{1}^{e}(L_{z}+m^{2})\right]\hat{R}_{0}^{e}\]
By the same arguments as for (\ref{egv:R1-o}), the solvability condition
applied to the expression above results in\begin{equation}
\beta_{1}^{e}=-\frac{(l-1)(l+2)(l^{2}-m^{2})((l+1)^{2}-m^{2})(3l(l+1)-2m^{2})}{6(l(l+1)-m^{2})^{2}},\label{eq:bt1-e}\end{equation}
 which again coincides with the corresponding result of \citet{Gail66}.

\subsection{\label{sub:visc-damp}Weak viscous damping}

There are three effects due to viscosity in this problem. Firstly,
viscosity appears in the normal stress balance condition (\ref{eq:nblnc-nd})
as a $O(\Ca)$ correction to the inviscid solution obtained above.
Secondly, viscosity also appears as a small parameter $\Ca$ in (\ref{eq:phip})
which again implies the same order correction when the leading-order
inviscid solution is substituted into this term. Thirdly, viscosity
enters the problem implicitly through the free-slip boundary condition
(\ref{eq:tblnc}) which was ignored by the inviscid solution but needs
to be satisfied when viscosity is taken into account. To satisfy this
condition, the leading-order solution needs to be corrected by the
viscous term in (\ref{eq:phip}), where $\Ca$ appears as a small
parameter at the higher-order derivative. For this small viscous term
to become comparable with the dominating magnetic term at the surface,
the expected correction has to vary over the characteristic length
scale $\delta\sim\sqrt{\Ca/\Cm}=\Ha^{-1},$ which is defined by the
Hartmann number $\Ha=B_{0}R_{0}\sqrt{\sigma/(\rho\nu)}$. Moreover,
for the viscous correction of the tangential velocity $\tilde{v}_{\tau}$
in the Hartmann layer to compensate for a $O(1)$ tangential stress
due to the leading-order inviscid solution, $\tilde{v}_{\tau}\sim\Ha^{-1}$
is required. Then the incompressibility constraint implies an associated
normal velocity component of an order in $\delta$ smaller than $\tilde{v}_{\tau},$
i.e., $\tilde{v}_{n}\sim\Ha^{-2}.$ This normal velocity correction
is subsequently negligible. But this not the case for the tangential
velocity correction $\tilde{v}_{\tau},$ which according to (\ref{eq:p})
is expected to produce a pressure correction $\tilde{p}\sim\Cm/\Ha^{2}\sim\Ca.$
The last is comparable with the normal viscous stress produced by
the leading-order inviscid flow. Taking into account the estimates
above and $\tilde{\varphi}\sim\delta\tilde{v}_{\phi}\sim\Ha^{-2},$
which follows from (\ref{eq:phi}), we search for a viscous correction
as \begin{eqnarray*}
\{\hat{p},\hat{\varphi},\hat{\vec{v}}\} & \sim & \{\hat{p}_{0},\hat{\varphi}_{0},\hat{\vec{v}}_{0}\}+\Ca\{\hat{p}_{01},\hat{\varphi}_{01},\hat{\vec{v}}_{01}\}+\{\Ca\tilde{p},\Ha^{-2}\tilde{\varphi},\Ha^{-1}\tilde{\vec{v}}\}\cdots,\\
\{\hat{R},\beta\} & \sim & \{\hat{R}_{0},\beta_{0}\}+\Ca\{\hat{R}_{01},\beta_{01}\}+\cdots,\end{eqnarray*}
where the terms with the tilde account for a Hartmann layer solution
localised at the surface.

\subsection{Eigenvalue perturbation for longitudinal modes}

We start with the core region, where the additive boundary layer corrections
are supposed to vanish. The first-order viscous corrections for the
pressure and potential $\{\hat{p}_{01}^{o},\hat{\varphi}_{01}^{o}\}(r,z)=z\{\hat{p}_{01}^{-},\hat{\varphi}_{01}^{-}\}(r)$
are obtained similarly to the leading-order inviscid solution (\ref{eq:p0f0}).
Now, instead of the kinematic and electric boundary conditions (\ref{eq:R1-cnd})
and (\ref{eq:phi-cnd}) derived in the inviscid approximation, we
have to use the original ones (\ref{eq:kinc}) and (\ref{eq:jR})
containing the velocity, which again follows from the Navier-Stokes
equation (\ref{eq:NS-nd}) including the viscous term $\sim\Ca.$

For the longitudinal modes, described by the odd solutions, (\ref{eq:NS-nd})
yields \begin{eqnarray}
\beta_{0}^{o}\hat{w}_{01}^{o}+\beta_{01}^{o}\hat{w}_{0}^{o} & = & -\hat{p}_{01}^{-}+L_{r}\hat{w}_{0}^{o},\label{eq:w01-o}\\
\hat{\vec{u}}_{01}^{o} & = & \vec{e}_{z}\times\vec{D}\hat{\varphi}_{01}^{o},\label{eq:u01-o}\end{eqnarray}
where $\hat{w}$ and $\hat{\vec{u}}$ are the velocity components
parallel and perpendicular, respectively, to the field direction $\vec{e}_{z},$
and $\vec{D}\equiv e^{-im\phi}\vec{\nabla}e^{im\phi}$ is a spectral
counterpart of the nabla operator for the azimuthal mode $m.$ Since,
as shown above, both the potential and velocity perturbations in the
Hartmann layer are higher-order small quantities and, thus, negligible
with respect to the core perturbation, the electric boundary condition
(\ref{eq:jR}) can be applied at $R=1$ directly to the first-order
core solution as $\partial_{R}\hat{\varphi}_{01}^{o}=-\vec{e}_{R}\cdot\vec{e}_{z}\times\hat{\vec{u}}_{01}^{o}.$
Taking into account (\ref{eq:u01-o}), this yields $\hat{\varphi}_{01}^{-}\equiv0$
and, hence, $\hat{\vec{u}}_{01}^{o}\equiv0.$ Consequently, the first-order
velocity perturbation in the core for the odd modes is again purely
longitudinal. Then, the kinematic constraint (\ref{eq:kinc}) for
the leading- and first-order terms takes, respectively, the form\begin{eqnarray*}
z\hat{w}_{0}^{o} & = & \beta_{0}^{o}\hat{R}_{0}^{o},\\
z\hat{w}_{01}^{o} & = & \beta_{0}^{o}\hat{R}_{01}^{o}+\beta_{01}^{o}\hat{R}_{0}^{o}.\end{eqnarray*}
These expressions combined with (\ref{eq:w01-o}) result in\begin{equation}
\beta_{0}^{o}(\beta_{0}^{o}\hat{R}_{01}^{o}+2\beta_{01}^{o}\hat{R}_{0}^{o}-zL_{r}z^{-1}\hat{R}_{0}^{o})=-\hat{p}_{01}^{o},\label{eq:p01-o}\end{equation}
which defines the first-order core pressure perturbation at $R=1.$
In addition, we need also the Hartmann layer pressure correction which
according to the estimates above is of the same order of magnitude
as the core one. 

To resolve the Hartmann layer, we introduce a stretched coordinate
$\tilde{R}=(1-R)/\delta$ (\citealt{Hinch91}), where $\delta=\Ha^{-1}$
is the characteristic Hartmann layer thickness. In the Hartmann layer
variables, (\ref{eq:phip}) takes the form\begin{equation}
(\Cm^{-1}\beta_{0}+z^{2}-\partial_{\tilde{R}}^{2})\partial_{\tilde{R}}^{2}\{\tilde{p},\tilde{\varphi},\tilde{\vec{v}}\}=0.\label{eq:pfv-bnd}\end{equation}
For $\Cm\gg1$, the inertial term $\sim\Cm^{-1}$ is negligible in
(\ref{eq:pfv-bnd}) with respect to the magnetic one $\sim z^{2},$
except for $|z|\lesssim\Cm^{-1/2}.$ First, ignoring this term, which,
as shown below, gives a next-order small correction, the solution
of (\ref{eq:pfv-bnd}) vanishing outside the Hartmann layer can be
written as \begin{equation}
\{\tilde{p},\tilde{\varphi},\tilde{\vec{v}}\}=\{\tilde{p}^{s},\tilde{\varphi}^{s},\tilde{\vec{v}}^{s}\}(z)e^{-|z|\tilde{R}},\label{eq:pfv-s}\end{equation}
where the index $s$ denotes the surface distribution of the corresponding
quantity. Then the free-slip boundary condition (\ref{eq:tblnc})
results in\begin{eqnarray}
\tilde{v}_{\phi}^{s} & = & -|z|^{-1}(imr^{-1}\hat{v}_{0,R}+\partial_{R}(\hat{v}_{0,\phi}/R)),\label{eq:vs-phi}\\
\tilde{v}_{\theta}^{s} & = & -|z|^{-1}(\partial_{\theta}\hat{v}_{0,R}+\partial_{R}(\hat{v}_{0,\theta}/R)).\label{eq:vs-theta}\end{eqnarray}
For the longitudinal modes, defined by the odd solutions, the leading-order
inviscid velocity is purely axial \begin{equation}
\hat{\vec{v}}_{0}^{o}=\vec{e}_{z}\hat{w}_{0}(r)=-\vec{e}_{z}{\beta_{0}^{o}}^{-1}\hat{p}_{0}^{-}(r).\label{eq:v0-o}\end{equation}
Substituting this into (\ref{eq:vs-theta}) and taking into account
that the radial pressure distribution at the surface is related to
the radius perturbation by (\ref{eq:R0-o}), we obtain\begin{equation}
\tilde{v}_{\theta}^{s}=\beta_{0}\frac{r(z^{2}-r^{2})}{z|z|}\frac{d}{dz}\frac{\hat{R}_{0}^{o}}{z}.\label{eq:vs-theta-o}\end{equation}
Pressure is related to the velocity by (\ref{eq:p}), which in the
dimensionless form reads as $\vec{\nabla}^{2}p=\Cm\partial_{z}v_{z}.$
In the Hartmann layer variables, this equation takes the form \begin{equation}
\partial_{\tilde{R}}^{2}\tilde{p}=rz\partial_{\tilde{R}}\tilde{v}_{\theta}.\label{eq:p-bnd}\end{equation}
Substituting the general solutions for pressure and velocity given
by (\ref{eq:pfv-s}) into (\ref{eq:p-bnd}) and using (\ref{eq:vs-theta-o}),
we find\begin{equation}
\tilde{p}^{s}=-rz|z|^{-1}\tilde{v}_{\theta}^{s}.\label{eq:ps-o}\end{equation}
Substituting the normal component of viscous stress \[
-2\partial_{R}\hat{v}_{0,R}^{o}=2\beta_{0}^{o}r^{2}\frac{d}{dz}\frac{\hat{R}_{0}^{o}}{z}\]
together with the core and boundary layer pressure contributions defined
by (\ref{eq:p01-o}) and (\ref{eq:ps-o}) into the normal stress balance
condition (\ref{eq:nblnc-nd}), we finally obtain\begin{equation}
(L_{z}+2-{\beta_{0}^{o}}^{2})\hat{R}_{01}^{o}=\beta_{0}^{o}\left[2\beta_{01}^{o}\hat{R}_{0}^{o}-z^{-2}(L_{z}+m^{2}+2)\hat{R}_{0}^{o}-2(1-z^{-2})\frac{d}{dz}\frac{\hat{R}_{0}^{o}}{z}\right].\label{eq:R01-o}\end{equation}
The sought for viscous damping rate is obtained in the usual way by
applying the solvability condition to (\ref{eq:R01-o}) that after
some algebra results in\begin{eqnarray}
\beta_{01}^{o} & = & -(2l+1)\frac{(l-m)!}{(l+m)!}\int_{0}^{1}\left[\frac{1}{2}(l(l+1)-m^{2}-2)\frac{P_{l}^{m}(z)}{z}\right.\label{sol:bet01-o}\\
 &  & \left.-(z-z^{-1})\frac{d}{dz}\frac{P_{l}^{m}(z)}{z}\right]\frac{P_{l}^{m}(z)}{z}\, dz=-(2l+1)\left[\frac{1}{2}(l(l+1)-m^{2})-1-I_{l}^{m}\right],\nonumber \end{eqnarray}
 where \begin{eqnarray}
I_{l}^{m} & = & \frac{(l-m)!}{(l-m)!}\int_{0}^{1}\frac{P_{l}^{m}(z)}{z}(z-z^{-1})\frac{d}{dz}\frac{P_{l}^{m}(z)}{z}\, dz\nonumber \\
 & = & \frac{((l-1)^{2}-m^{2})I_{l-2}^{m}+(2l-1)(l(l-1)-m^{2})}{l^{2}-m^{2}}\label{eq:I-lm}\end{eqnarray}
can be calculated from the above recurrence relation starting with
$l=m+1$ and taking into account that $I_{l}^{m}=0$ for $l<m.$ For
the modes with $m=l-1$, we have $\beta_{01}^{o}=\frac{1}{2}(2l+1)(l-1),$
which is the half of the corresponding viscous damping rate without
the magnetic field (\citealt{Lamb93}). Although the viscous damping
rate increases for smaller $m,$ as seen from the numerical values
of $-\beta_{01}^{o}$ for the first 7 longitudinal oscillation modes
calculated by the Mathematica (\citealt{Wolf96}) and shown in table
\ref{tab:bet01-o}, it remains below its non-magnetic counterpart
up to $l=5$ modes.

Note that the r.h.s of (\ref{eq:R01-o}) has a simple pole $(z^{-1})$
singularity at $z=0,$ which is due to the neglected inertial term
in (\ref{eq:pfv-bnd}). As discussed above, this term becomes relevant
for $|z|\lesssim\Cm^{-1/2},$ where it cuts off the singularity at
$z^{-1}\sim\Cm^{1/2}.$ This cut-off integrated in (\ref{sol:bet01-o})
over $|z|\lesssim\Cm^{-1/2},$ where $P_{l}^{m}(z)\sim z$ for the
odd modes, results in the damping rate correction $O(\Cm^{-1/2}),$
which is a higher-order small quantity. 

\begin{center}
\begin{table}
\begin{centering}
\begin{tabular}{cccccccc}
 & $m=0$ & 1 & 2 & 3 & 4 & 5 & 6\tabularnewline
\hline
$l=2$ &  & $\frac{5}{2}$ &  &  &  &  & \tabularnewline
3 & $\frac{35}{3}$ &  & $7$ &  &  &  & \tabularnewline
4 &  & $\frac{51}{2}$ &  & $\frac{27}{2}$ &  &  & \tabularnewline
5 & $\frac{154}{3}$ &  & $44$ &  & $22$ &  & \tabularnewline
6 &  & $\frac{169}{2}$ &  & $\frac{403}{6}$ &  & $\frac{65}{2}$ & \tabularnewline
7 & $135$ &  & $125$ &  & $95$ &  & $45$\tabularnewline
\end{tabular}
\par\end{centering}

\caption{\label{tab:bet01-o}The viscous damping rates $-\beta_{01}^{o}$ for
the first 6 longitudinal oscillation modes.}

\end{table}

\par\end{center}

\subsection{Viscous energy dissipation}

Viscous damping rate can be found in an alternative much simpler way
by considering the energy balance following from the dot product of
(\ref{eq:NS-nd}) and $\vec{v},$ which integrated over the drop volume
yields \begin{equation}
\frac{1}{2}\partial_{t}\int_{V}\vec{v}^{2}\, dV+\int_{S}(\vec{\nabla}\cdot\vec{n})\vec{v}\cdot d\vec{s}=-\int_{V}(2\Ca\breve{\varepsilon}^{2}+\Cm\vec{j}^{2})\, dV,\label{eq:nrg}\end{equation}
where the first and second term on the l.h.s. stand for the time-variation
of kinetic and surface energies, while the terms on the r.h.s. with
the rate-of-strain tensor $(\breve{\varepsilon})_{i,j}=\frac{1}{2}(v_{i,j}+v_{j,i})$
and the dimensionless current density $\vec{j}=-\vec{\nabla}\varphi+\vec{v}\times\vec{\epsilon}$
account for the viscous and ohmic dissipations, respectively. As estimated
above, viscosity gives rise to the tangential current density $\sim\Ha^{-1}$
in the Hartmann layer of the thickness $\sim\Ha^{-1}$ that according
to (\ref{eq:nrg}) produces the ohmic dissipation $\sim\Cm/\Ha^{3}\sim\Ca/\Ha$
which for $\Ha\gg1$ is negligible with respect to the viscous dissipation
$\sim\Ca.$ Note that although the contribution of the Hartmann layer
to the normal stress balance is important, its contribution to the
energy dissipation is still negligible. This fact results in a substantial
simplification of the solution procedure for the viscous damping rate.

Thus, neglecting the ohmic dissipation and averaging the rest of (\ref{eq:nrg})
over the period of oscillation and taking into account that the mean
kinetic and surface energies for small amplitude harmonic oscillations
are equal, we obtain a simple expression for the viscous damping rate
in terms of inviscid leading order solution (\citealt{Landau87})
\begin{equation}
\beta_{01}=-\int_{V}|\breve{\hat{\varepsilon}}_{0}|^{2}\, dV/\int_{V}|\vec{\hat{v}}_{0}|^{2}\, dV.\label{eq:bet01}\end{equation}
For the longitudinal modes, this equation takes the form \begin{equation}
\beta_{01}^{e}=-\frac{\int_{0}^{1}[(rz^{-1}\partial_{z}\hat{w}_{0}^{o})^{2}+(m\hat{w}_{0}^{o}/r)^{2}]z^{2}\, dz}{\int_{0}^{1}\left.\hat{w}_{0}^{o}\right.^{2}z^{2}\, dz}.\label{eq:bt01-o}\end{equation}
Substituting $\hat{w}_{0}^{o}(z)=\beta_{0}^{o}R_{0}^{o}P_{l}^{m}(z)/z$
from (\ref{eq:w0-o}) into (\ref{eq:bt01-o}), after some algebra
the last can be shown to be equivalent to (\ref{sol:bet01-o}).

This approach is particularly useful for the transversal modes for
which the conventional eigenvalue perturbation solution becomes excessively
complicated and, thus, it is omitted here. In this case, using (\ref{sol:v0-e})
we can represent (\ref{eq:bet01}) in terms of scalar potential \begin{equation}
\beta_{01}^{e}=-\frac{\int_{0}^{1}\left[\left(r\partial_{r}(r^{-1}\partial_{r}\hat{\varphi}_{0}^{e})+m^{2}\hat{\varphi}_{0}^{e}/r\right)^{2}+\left(2m\partial_{r}(\hat{\varphi}_{0}^{e}/r)\right)^{2}\right]z^{2}\, dz}{\int_{0}^{1}\left[(\partial_{r}\hat{\varphi}_{0}^{e})^{2}+(m\hat{\varphi}_{0}^{e}/r)^{2}\right]z^{2}\, dz}.\label{eq:bt01-e}\end{equation}
Substituting $\hat{\varphi}_{0}^{e}(z)=im^{-1}\beta_{0}^{e}R_{0}^{e}P_{l}^{m}(z)$
from (\ref{sol:phi0-p}) into (\ref{eq:bt01-e}), after a lengthy
algebra we obtain\begin{equation}
\beta_{01}^{e}=-(2l+1)\frac{l(l+1)(l-2)-m^{2}(l-3)+(l^{2}-m^{2})I_{l-1}^{m}}{2(l(l+1)-m^{2})},\label{sol:bt01-e}\end{equation}
where $I_{l-1}^{m}$ is defined by (\ref{eq:I-lm}). Note that for
2D modes, defined by $m=l,$ which are not affected by the magnetic
field, we recover the well-known non-magnetic result $\beta_{01}^{e}=-(2l+1)(l-1)$
(\citealt{Lamb93}). For other indices, (\ref{sol:bt01-e}) can be
verified by a direct integration of (\ref{eq:bt01-e}) using the Mathematica
(\citealt{Wolf96}). As seen from the numerical values shown in table
\ref{tab:bet01-e}, the next even mode with $m=l-2$ has the viscous
damping rate which is by the factor of $(l-2)/(l-4/5)$ lower than
the non-magnetic counterpart given by $m=l$. Only for the modes with
$m\leq l-4,$ the viscous damping rate in the magnetic field becomes
higher than that without the field.

\begin{center}
\begin{table}
\begin{centering}
\begin{tabular}{cccccccc}
 & $m=1$ & 2 & 3 & 4 & 5 & 6 & 7\tabularnewline
\hline
$l=2$ &  & $5$ &  &  &  &  & \tabularnewline
3 & $\frac{70}{11}$ &  & $14$ &  &  &  & \tabularnewline
4 &  & $\frac{135}{8}$ &  & $27$ &  &  & \tabularnewline
5 & $\frac{1232}{29}$ &  & $\frac{220}{7}$ &  & $44$ &  & \tabularnewline
6 &  & $\frac{1339}{19}$ &  & $50$ &  & $65$ & \tabularnewline
7 & $\frac{1350}{11}$ &  & $\frac{4930}{47}$ &  & $\frac{2250}{31}$ &  & 90\tabularnewline
\end{tabular}
\par\end{centering}

\caption{\label{tab:bet01-e}The viscous damping rates $-\beta_{01}^{e}$ for
the first 6 transversal oscillation modes.}

\end{table}

\par\end{center}

The approach above is not directly applicable to the axisymmetric
transversal modes which, as discussed at the end of $\S$\ref{sub:T-freq},
are stationary in the leading-order inviscid approximation. For these
overdamped modes, a flow with the velocity $\sim1/\Cm$ relative to
the leading-order radius perturbation appears only in the first-order
approximation, which according to (\ref{eq:nrg}) produces the same
order ohmic dissipation. In this case, dissipation takes place on
the account of the surface energy reduction, while that of the kinetic
energy is negligible because it is by $\sim1/\Cm^{2}$ smaller than
the former. The contribution of the viscous dissipation in (\ref{eq:nrg})
is $\sim\Ca/\Cm^{2},$ which for a low viscosity and a high magnetic
field is much smaller than the ohmic dissipation $\sim1/\Cm,$ and,
thus negligible with respect to the latter.

\section{\label{sec:conc}Conclusion}

In the present study, we have considered small-amplitude oscillations
of a conducting liquid drop in a uniform DC magnetic field. Viscosity
was assumed to be small but the magnetic field strong. Combining the
regular and matched asymptotic expansion techniques we obtained a
relatively simple solution to the associated eigenvalue problem. Firstly,
we showed that the eigenmodes of shape oscillations are not affected
by strong magnetic field. Namely, they remain the spherical harmonics
as in the non-magnetic case. Strong magnetic field, however, constrains
the liquid flow associated with the oscillations and, thus, reduces
the oscillations frequency by increasing apparent inertia of the liquid.
In such a field, liquid oscillates in a two-dimensional (2D) way as
solid columns aligned with the field. Two types of oscillations are
possible: longitudinal and transversal to the field. Such oscillations
are weakly damped by strong magnetic field -- the stronger the field,
the weaker the damping, except for the axisymmetric transversal and
2D modes. The former are magnetically overdamped because the incompressibility
constraint does not permit an axially uniform radial flow. The latter,
which are transversal modes defined by the spherical harmonics with
equal degree and order, $l=m$, are not affected by the magnetic field
because these modes are naturally invariant along the field. In a
uniform magnetic field, no electric current is induced and, thus,
no electromagnetic force acts on such a 2D transversal flow because
the associated e.m.f. is irrotational. Because the magnetic damping
for all other modes decreases inversely with the square of the field
strength, the viscous damping may become important in a sufficiently
strong magnetic field. Consequently, the relaxation of axisymmetric
transversal modes, whose viscous damping is negligible relative to
the magnetic one, can be used to determine the electrical conductivity,
while the damping of $l=m$ modes can be used to determine the viscosity.
The damping of all other modes is affected by both the viscous and
ohmic dissipations. Although the latter reduces inversely with the
square of the field strength while the former stays constant, an extremely
strong magnetic field may be required for the viscous dissipation
to be become dominant.

As an example, let us consider a drop of Nickel of $1\, cm$ in diameter
$(R_{0}=5\times10^{-3}\, m)$ which, at the melting point $(1455^{\circ}C),$
has the surface tension $\gamma=1.8\, N/m,$ density $\rho=7.9\times10^{3}\, kg/m^{3},$
the dynamic viscosity $\eta=4.9\times10^{-3}\, Ns/m^{2}$ and the
electrical conductivity $\sigma=1.2\times10^{6}\, S/m$ (\citealt{Smithells}).
The capillary time scale and frequency of non-magnetic fundamental
mode $(l=2)$ for such a drop are $\tau_{0}=\sqrt{R_{0}^{3}\rho/\gamma}\approx23\, ms$
and $f=\sqrt{l(l-1)(l+2)}/(2\pi\tau_{0})\approx19\, Hz,$ respectively.
The viscous damping time without the magnetic field (\citealt{Lamb93})
is $\tau_{v}/((2l+1)(l-1))\approx8\, s$, where $\tau_{v}=\rho R_{0}^{2}/\eta\approx40\, s$
is the viscous time scale. Note that weak-viscosity approximation
is applicable in this case because $\Ca=\tau_{0}/\tau_{v}=5.8\times10^{-4}$
is small. In the magnetic field of $B=5\, T,$ for which $\Cm=\sigma B^{2}R_{0}^{2}/\sqrt{\rho\gamma R_{0}}\approx87\gg1,$
the oscillation frequency of longitudinal fundamental mode $(l,m)=(2,1)$
drops according to equation (\ref{sol:bt0-o}) to $f_{2,1}^{o}=\sqrt{(l-1)(l+2)}/(2\pi\tau_{0})\approx14\, Hz.$
The corresponding viscous damping time increases by the factor of
two to $-\tau_{\nu}/\beta_{01}^{o}\approx16\, s,$ where $-\beta_{01}^{o}=5/2$
according to table \ref{tab:bet01-o}. The magnetic damping time of
this mode, for which (\ref{eq:bt1-o}) yields $\beta_{1}^{o}=-4,$
is $-\tau_{0}\Cm/\beta_{1}^{o}\approx0.5\, s.$ According to this
formula, for the magnetic damping time to exceed the viscous one,
a magnetic field of $B\gtrsim20\, T$ is necessary. The relaxation
time for the axisymmetric fundamental mode $(l,m)=(2,0),$ which is
magnetically over-damped, is $-\tau_{0}\Cm/\beta_{1}^{e}\approx28\, ms,$
where $\beta_{1}^{e}=72$ follows from (\ref{eq:bt1-e}). The magnetic
field affects neither the frequency nor the damping rate of $(l,m)=(2,2)$
transversal oscillation mode, which is naturally invariant along the
field. For the same reason, there is no magnetic damping of this mode
either. The first oscillatory transversal mode is $(l,m)=(3,1)$ whose
frequency drops according to (\ref{sol:bt0-e}) from $f_{l}=\sqrt{l(l-1)(l+2)}/(2\pi\tau_{0})\approx38\, Hz$
without the magnetic field to $f_{3,1}^{e}=\sqrt{\frac{(l-1)(l+2)}{l(l+1)-m^{2}}}/(2\pi\tau_{0})\approx5\, Hz$
in a strong magnetic field. The magnetic damping time for this mode
in a $5\, T$ magnetic field is $-\tau_{0}\Cm/\beta_{1}^{e}\approx36\, ms,$
where $\beta_{1}^{e}=6800/121$ follows from (\ref{eq:bt1-e}). The
viscous damping time for this mode is $-\tau_{\nu}/\beta_{01}^{e}\approx6\, s,$
where $\beta_{01}^{e}=70/11$ follows from table \ref{tab:bet01-e}.
The viscous damping is small relative to the magnetic one for this
mode, and a magnetic field of about $65\, T$ would be necessary for
the magnetic damping time to become as long as the viscous one.

In conclusion, this theoretical model provides a basis for the development
of new measurement method of surface tension, viscosity and electrical
conductivity of liquid metals using oscillating drop technique in
a strong superimposed DC magnetic field.

\begin{acknowledgements}

The author would like to thank Agris Gailitis and Ra\'ul Avalos-Z\'u\~niga
for constructive comments and stimulating discussions.

\end{acknowledgements}

\end{document}